\documentstyle[twocolumn,floats,aps]{revtex}
\begin{document}

%\draft
\twocolumn[\hsize\textwidth\columnwidth\hsize\csname
@twocolumnfalse\endcsname

\title{Learning properties of Support Vector Machines}
\author{Arnaud Buhot \and Mirta B. Gordon\cite{cnrs}}
\address{D\'epartement de Recherche Fondamentale sur la Mati\`ere
Condens\'ee,\\ CEA/Grenoble, 17 rue des Martyrs, 38054 Grenoble Cedex 9,
France}
\date{February 16, 1998}

\maketitle

\begin{abstract}

In this article, we study the typical learning properties of the recently proposed Support Vectors Machines. The generalization error on linearly separable tasks, the capacity, the typical number of Support Vectors, the margin, and the robustness or noise tolerance of a class of Support Vector Machines are determined in the framework of Statistical Mechanics. The robustness is shown to be closely related to the generalization properties of these machines.
 
\end{abstract}

]

{\it Support Vector Machines}, recently proposed  
to solve the problem of learning classification tasks from examples, have  aroused a great deal of interest due to the simplicity of their implementation, and to their remarkable performances on difficult tasks~\cite{Vapnik,Burges}. Classification of data is a very general problem, as  many real-life applications, 
like pattern recognition, medical diagnosis, etc., may be 
cast as classification tasks. In the last few years, much work has 
been done to understand how high-performance learning may be achieved, mainly 
within the paradigm of neural networks. These are systems composed 
of interconnected neurons, which are two-state units like spins. The neuron's  
state is determined, like in magnets, by the sign of the weighted 
sum of its inputs, which acts as an external field, and of the states of its neighbors. Learning with neural networks 
means determining their connectivity and the weights of the connections. The aim is to classify correctly not only the examples, or 
training patterns, but also new data, as we expect that the learning system will be able to generalize. A single neuron connected 
to its inputs, the {\it simple perceptron} (SP), is the elementary  
neural network. It separates the input patterns in two classes by 
a hyperplane orthogonal to a vector whose components 
are the connection weights. Thus, the SP can learn without errors only 
Linearly Separable (LS) tasks. Most classification problems are 
not LS, requiring learning machines with more degrees 
of freedom. However, the relationship between the machine's complexity,
its learning capacity and its generalization ability is still an open problem. 
Feedforward layered networks, the {\it multilayered 
perceptrons}, are the most popular learning machines. Their 
architecture is usually found through a trial and error procedure, 
in which the weights are determined with Backpropagation~\cite{Backprop}, a 
learning algorithm that performs a gradient descent on a cost 
function. Its main drawback is that it usually gets trapped in 
metastable states. Growth heuristics that avoid using 
Backpropagation have also been proposed~\cite{Torres_Gordon}.

Support Vector Machines (SVM)~\cite{Vapnik,Burges,Cortes_Vapnik} are an alternative solution to the learning problem, whose {\it typical} properties have not been studied theoretically yet. 
The idea underlying SVM is to map the patterns from 
the input space to a new space, the {\it feature-space}, through 
a non-linear transformation chosen {\it a priori}. Provided 
that the dimension of the feature-space is large enough, the 
image of the training patterns will be LS, {\it i.e.} learnable 
by a SP. It is well known that, if the training set 
is LS, there is an infinite number of error-free separating 
hyperplanes. Among them, the {\it Maximal Stability Perceptron} 
(MSP) has weights that maximize the distance of the patterns closest to it. 
The SVM weight vector is that of the MSP in feature-space. The patterns closest to the separating hyperplane are called {\it Support Vectors} (SV)~; their distance to the hyperplane is the {\it maximal stability} or SV-margin. 
The important point is that the SV determine uniquely the MSP. Their number is proportional to the number of training patterns, and {\it not} to the dimension of the feature-space (which may be huge). Thus, 
increasing the feature-space dimension does not necessarily 
increase the number of parameters to be learned, a fact that makes the SVM 
very attractive for applications. For example, in the problem of 
digit recognition~\cite{Vapnik}, the input space of dimension 
$256$ needs to be mapped onto a space of dimension $256^7 \sim  
10^{16}$, but the number of parameters to be determined is as 
low as $422$. However, in spite of the high performance reached by SVMs in realistic problems~\cite{Burges}, a theoretical understanding of their properties is still lacking. 

We consider,  within the framework of Statistical Mechanics, 
SVMs defined by particular families of mappings between the input-space and the feature-space. We address several important questions about these machines. The generalization error in the particular
case of learning a LS task is shown to decrease slower than that of a SP (in input-space) as a function of the number of training patterns. The capacity increases 
proportionally to the dimension of the feature-space. The number of SV and the SV-margin present interesting scaling with the number of features. The probability of misclassification of training patterns corrupted after learning is shown to be a decreasing function of the SV-margin. This property, that we call robustness or noise-tolerance, may account for the good generalization performance of SVMs in applications.

We assume that we are given a training set of $P$ independent 
$N$-dimensional vectors, the {\it training
patterns} $\left\{\bbox{\xi}^{\mu} \right\}_{\mu = 1,\cdots,P}$, 
and their
corresponding classes $\tau^{\mu} = \pm 1$. The patterns are
supposed to be drawn with a probability density
$P({\bbox \xi}) = (2\pi)^{-N/2} \exp \left( - {\bbox \xi}^2/2
\right)$, and the classes $\tau$ are given by an unknown function 
$\tau(\bbox{\xi})$ called supervisor or teacher. 

We focus on SVMs defined by a nonlinear transformation $\Phi$ that maps the $N$-dimensional input space to a $(k + 1) 
N$-dimensional feature-space through

\begin{equation}
\bbox{\xi} \rightarrow \Phi \left( \bbox{\xi} \right) = 
\left\{ \bbox{\xi} , \phi(\lambda_{1}) \bbox{\xi}, \cdots,
\phi(\lambda_{k}) \bbox{\xi} \right\},
\end{equation}

\noindent where the $\lambda_{i}$ are functions of $\bbox{\xi}$. 
The components $\phi(\lambda_{i}) \bbox{\xi}$ 
($i=1,\cdots,k$) are the {\it new features} that hopefully should 
make the task linearly separable in feature-space. 

In the following we consider odd functions $\phi$, and 
$\lambda_{i} = \bbox{\xi} \cdot {\bf B}_{i}$ where 
$\left\{ {\bf B}_{i} \right\}_{i = 1,\cdots,k}$ is a set of $k$ unitary orthogonal vectors (${\bf B}_i \cdot {\bf B}_j = 
\delta_{ij}$). With this choice, the new features are 
uncorrelated. For example, the $k$ first generators $\left\{ e_1, e_2, 
\cdots, e_k \right\}$ of the input space ($e_1 = (1,0,\cdots,0)$, $e_2 = 
(0,1,0,\cdots,0)$, $\cdots$) are one possible realization of the ${\bf B}_{i}$. In the thermodynamic limit considered below, any set of $k$ randomly selected normalized vectors ${\bf B}_i$ satisfies the orthogonality constraint with probability one. The functions $\phi(\lambda) = {\rm sign}
(\lambda)$ and  $\phi(\lambda) = \lambda$ are of particular interest. If $k = N$, a SVM using the latter can implement all the possible 
discriminating surfaces of second order in input space. More complicated 
transformations $\Phi$, equivalent to higher order surfaces, may be 
considered (for examples, see~\cite{Vapnik}).

The output of the SVM to a pattern ${\bbox \xi}$ is $\sigma = {\rm sign} 
\left ( {\bf J} \cdot \Phi({\bbox \xi}) \right)$, where ${\bf J} = 
\left\{ {\bf J}_0, {\bf J}_1, \cdots, {\bf J}_k \right\}$ is a 
$(k+1)N$-dimensional vector. Hereafter we consider normalized weights, ${\bf J} \cdot {\bf J} = (k+1)N$ without any lack of generality, but we {\it do not} impose any constraint to the normalization of each 
$N$-dimensional vector ${\bf J}_i$. The stability of a training pattern 
$\bbox{\xi}^{\mu}$ of class $\tau^{\mu}$ in feature-space is 

\begin{equation}
\label{eq:stab}
\gamma^{\mu} = \frac{\tau^{\mu} \ {\bf J} \cdot
\Phi(\bbox{\xi}^{\mu})}{\sqrt{(k+1)N}}.
\end{equation}

\noindent Geometrically, $|\gamma^{\mu}|$ is the distance of the image $\Phi(\bbox{\xi}^{\mu})$ of pattern $\bbox{\xi}^{\mu}$ 
to the hyperplane orthogonal to ${\bf J}$. The aim of learning is to determine a vector ${\bf J}$ such that $\sigma^{\mu} = \tau^{\mu}$, or equivalently 
$\gamma^{\mu} > 0$, for all $\mu$. Any vector ${\bf J}$ that meets these learning conditions separates linearly, in the feature-space, the image $\Phi$ of patterns with output $+1$ from those with output $-1$. Due to the
non-linearity of $\Phi$, this  separation is not linear in input
space. More general SVMs, 
that use a Kernel $K({\bf J},\Phi(\bbox{\xi}))$ instead of the inner 
product in Eq.(\ref{eq:stab}), have been proposed~\cite{Vapnik}, but we 
restrict to the inner product in the following.

The SV-margin is

\begin{equation}
\kappa_{\rm max} ( {\bf J^*} ) = \max_{\bf J} \inf_{\mu} \gamma^{\mu},
\end{equation}

\noindent where ${\bf J^*}$, the MSP weight vector in feature-space, 
is a linear combination of the SV~\cite{Vapnik,Gerl_Krey}, ${\bf J^*} = 
\sum_{\mu \in SV} x^{\mu} \tau^{\mu} \Phi({\bbox \xi}^{\mu})$. The 
$x^{\mu}$ are positive parameters to be determined
by the learning algorithm, which has to determine also the number of SV. Generally, this number is small 
compared with the feature-space dimension, a fact that 
allows to increase the latter considerably without
increasing dramatically the number of parameters to be determined. The SVM in input-space ($k=0$) or {\it linear SVM} is the usual MSP, whose properties have extensively been studied (see~\cite{Gordon_Grempel} and references therein). 

We obtain the generic properties of the SVM through the 
by now standard replica approach~\cite{Opper_Kinzel}. Results are obtained in 
the thermodynamic limit, in which the input space dimension and 
the number of training patterns go to infinity ($N \rightarrow + 
\infty$, $P \rightarrow +\infty$) keeping the reduced number of 
patterns $\alpha = P/N$ constant. In this limit, the SVM properties 
are independent of the training set. The appropriate cost function 
is $E({\bf J}, {\cal L}_{\alpha},\kappa) = \sum_{\mu} \Theta (\kappa 
- \gamma^{\mu})$, where $\Theta$ is the Heaviside function and 
${\cal L}_{\alpha}$ represents the training set. It  
counts the number of training patterns that have a stability smaller 
than $\kappa$ in feature-space. The  largest 
value of $\kappa$ that satisfies $E({\bf J^*},{\cal L}_{\alpha},\kappa) 
= 0$ is the SV-margin. The weight vector ${\bf J^*}$ defines the SVM. Its generic properties are determined by the free energy

\begin{equation}
\label{eq:free_energy}
f = \lim_{N \rightarrow +\infty} \lim_{\beta \rightarrow +\infty} - 
\frac{1}{\beta N} \left< \ln Z \right>,
\end{equation}

\noindent where $Z = \int dP({\bf J}) \exp \left( - \beta E({\bf J},
{\cal L}_{\alpha}, \kappa) \right)$ is the partition function,
$dP({\bf J}) = d{\bf J} \  \delta {\bf (} (k+1)N - {\bf J} \cdot
{\bf J} {\bf )}$ and $\beta$ is an inverse temperature. In Eq.
(\ref{eq:free_energy}), the bracket stands for the average over all
the possible training sets ${\cal L}_{\alpha}$ at given $\alpha$. If
the problem is LS, then $f=0$ for $\kappa \ge 0$, meaning that error-free learning is possible. But in general, the probability of error-free learning vanishes beyond some value of $\kappa$. The maximal value of $\kappa$ for which $f=0$ is the {\it typical} value of $\kappa_{\rm max}(k,\alpha)$. The free energy is calculated using the replica trick
$\left< \ln Z \right> = \lim_{n \rightarrow 0} \ln \left< Z^n \right>/n$.

We consider first the case of a teacher that is a SP in input space, of (unknown) $N$-dimensional normalized weigths ${\bf K}$ (${\bf K} \cdot {\bf K} = 1$). Thus, the classes of the 
training patterns ${\bbox \xi}^{\mu}$ are $\tau^{\mu} = {\rm sign} 
\left( {\bf K} \cdot {\bbox \xi}^{\mu} \right)$. In this case, 
an error-free solution exists for all $\alpha$, and we are interested in the generalization error $\epsilon_g(k,\alpha)$, which is the
probability that the trained SVM misclassifies a new pattern
${\bbox \xi}$. Clearly, we do not expect that a SVM with $k > 0$ will perform well on this task, as it corresponds to a case where the {\it a priori} selected feature-space is too complex. However, this may well be the case in real applications. We begin by considering this LS problem mainly because other properties considered below, like the capacity and robustness, can easily be deduced by disregarding, or setting to zero, some of the order parameters introduced here. These are,

\begin{mathletters}
\label{order_param}
\begin{eqnarray}
\label{order_param_a}
R^a & = & \frac{{\bf J}_0^a \cdot {\bf K}}{\sqrt{{\bf J}_0^a \cdot {\bf J}_0^a}},\\
\label{order_param_b}
v_i^a & = & \frac{{\bf J}_i^a \cdot {\bf J}_i^a}{N}, \\
\label{order_param_c}
c_i^{ab} & = & \lim_{\beta \, \rightarrow +\infty} \beta \frac{({\bf J}_i^a 
- {\bf J}_i^b)^2}{2N} \;\;\;\;\;\; (a \neq b),
\end{eqnarray}
\end{mathletters}

\noindent for $i = 0, \cdots, k$. ${\bf J}^a$ and ${\bf J}^b$ are the weight vectors of replicas $a$ and $b$. The cross-overlaps ${\bf J}_i^a\cdot {\bf J}_j^b$ ($i \neq j$), and ${\bf K} \cdot {\bf B}_i$ may be neglected for $k \ll N$, as they are of order $1/\sqrt{N}$. The parameters $c_i^{ab}$ are 
a generalization of the parameter $x^{ab} = \lim_{\beta \rightarrow +\infty} 
\beta (1 - q^{ab})$ in~\cite{Gardner_Derrida}. In fact, Gardner and 
Derrida~\cite{Gardner_Derrida,Gardner} considered a single perceptron ($k=0$) with normalized weights ${\bf J}_0$ (${\bf J}_0 \cdot {\bf J}_0 = N$), so that  
$({\bf J}_0^a - {\bf J}_0^b)^2/2N = 1 - {\bf J}_0^a 
\cdot {\bf J}_0^b/N = 1 - q^{ab}$ in their notations. We assume replica symmetry, {\it i.e.} $R^a=R$, $v_i^a=v_i$, $c_i^{ab}=c_i$ for all $a,b$. The parameter $R$ represents trivially the overlap between the first $N$ 
components of vector ${\bf J}$ with the teacher ${\bf K}$. The overlap 
between ${\bf J}_i$ and ${\bf K}$ may be neglected for $i \ge 1$, since 
for odd functions $\phi$ and uncorrelated vectors ${\bf B}_i$ the 
new features are uncorrelated. If $\phi$ were even, this would not be the case. The parameters $v_i$ are proportional to the norm of the ${\bf J}_i$. The sense of the parameters $c_i$ 
is more involved. They reflect how fast the fluctuations of ${\bf J}_i$  around the minimum of the cost function decrease as the temperature vanishes ($\beta \rightarrow +\infty$). In the case of a degenerate continuum of minima, these fluctuations decrease too slowly, and the $c_i$ diverge. This is the case for $\kappa < \kappa_{\rm max}$. 

A symmetry between the $k$ vectors ${\bf J}_i$, $i \ge 1$, due to the 
invariance with respect to permutations of the ${\bf B}_i$, together with the fact that the ${\bf B}_i$ are uncorrelated with ${\bf K}$, allows 
to take $v_i = v_1$ and $c_i = c_1$ for $i \ge 1$. Introducing $\tilde v_1 = v_1/v_0$, where $v_0$ is determined by the normalization condition ${\bf J} \cdot {\bf J}/N = k + 1 = v_0 + k \tilde v_1 v_0$, $\tilde c_1 = c_1/c_0$
and $\tilde c_0 = c_0/(1+k)$, the free energy is $f(k, \alpha, \kappa) = \max_{\tilde v_1,\tilde c_1,\tilde c_0} \min_{R} g(k, \alpha, \kappa; \tilde v_1,\tilde c_1,\tilde c_0, R)$, with

\begin{eqnarray}
\label{free_energy}
\nonumber & g(k, \alpha, \kappa; \tilde v_1,\tilde c_1,\tilde c_0, R)  = 
- \frac{\tilde c_1 \, (1 - R^2) + k \tilde v_1}
{2 \, \tilde c_0 \, \tilde c_1 \, ( 1 + k \tilde v_1 )}  \\ 
\nonumber  &   \, + \,  \frac{\alpha}{\tilde c_0} \int D\lambda_1 \cdots \int D\lambda_k 
\int_{\kappa a - b}^{\kappa a} \ Dy \ \frac{(\kappa -y/a)^2}{e} H \left( 
-\frac{y R}{\sqrt{e}} \right) \\
 &   + \, 2 \, \alpha  \int D\lambda_1 \cdots \int D\lambda_k  
\int_{-\infty}^{\kappa a - b} \ Dy H \left( -\frac{y R}{\sqrt{e}} 
\right).
\end{eqnarray}

\noindent $Dy = dy \exp (-y^2/2) /\sqrt{2\pi}$, 
$H(x) = \int_{x}^{+\infty} Dy$, and $a$, $b$, $e$ stand for

\begin{mathletters}
\label{parameters}
\begin{eqnarray}
\label{parameter_a}
a & = & \sqrt{\frac{1 + k \tilde v_1}{e+R^2}}, \\
b & = & a \left[ \tilde c_0 \, \left( 1 + \tilde c_1 \sum_{i=1}^{k} \phi^2(\lambda_i)\right) \right]^{1/2}, \\
\label{parameter_c}
e & = & 1 - R^2 + \tilde v_1 \sum_{i=1}^k \phi^2(\lambda_i).
\end{eqnarray}
\end{mathletters}

The generalization error $\epsilon_g(k, \alpha)$ writes

\begin{equation}
\label{gen_error}
\epsilon_g(k, \alpha) = \frac{1}{\pi} \int D\lambda_1 \cdots \int D\lambda_k 
\, \arccos \left( \frac{R}{\sqrt{e+R^2}} \right),
\end{equation}

\noindent where $R$ and $e$ extremize $g(k, \alpha, \kappa; \tilde v_1,\tilde c_1,\tilde c_0, R)$. In particular, the maximal stability $\kappa_{\rm max}(k, \alpha)$ is the 
largest value of $\kappa$ that satisfies $\tilde c_0(\alpha,\kappa) = +\infty$ 
since $f$ is non zero for finite values of $\tilde c_0$. 

If $\phi(\lambda) = {\rm sign} (\lambda)$, the extremization of (\ref{free_energy})  with respect to $\tilde v_1$ 
and $\tilde c_1$ gives $\tilde v_1 = 1 - R^2$ and $\tilde c_1 = 1$. 
Notice that for $R=1$ (which corresponds to $\alpha = \infty$), 
$\tilde v_1=0$ (thus, $v_1=0$) as expected: the new features are 
irrelevant because the task is LS. The fact that  $\tilde c_1 = 1$ 
means that the fluctuations of ${\bf J}_0$ and ${\bf J}_i$, $i \ge 1$, 
have the same behaviour in the limit $\beta \rightarrow \infty$ despite 
the fact that their norms are different ($\tilde v_1 \ne 1$). After 
introduction of these values for $\tilde v_1$ and $\tilde c_1$ in 
(\ref {free_energy}), we obtain $g(k, \alpha, \kappa; \tilde c_0, R)  
=  g \left(0, \alpha/(k+1), \kappa; \tilde c_0, R/\sqrt{1 + k (1- R^2)} \right)$, where the right hand side term corresponds to a SP 
trained with a training set of reduced size $\alpha/(k+1)$ 
having an  overlap $R/\sqrt{1 + k (1- R^2)}$ with the teacher. 
After introducing these values of the order parameters in 
(\ref{parameter_c}) and (\ref{gen_error}), we obtain $\epsilon_g(k,\alpha) = \epsilon_g(0,\alpha/(k+1))$. As expected, the generalization error of the SVM with $k>0$ on a LS task is larger than the one of the linear SVM. This is due to an entropic effect, as the SVM's phase space grows with $k$ whereas the size of 
the space of functions considered, limited to the LS ones, remains 
the same. For large $\alpha$, the generalization error vanishes as $0.5005 
\, (k+1)/\alpha$, to be compared to the linear SVM that has $\epsilon_g \sim 0.5005/\alpha$~\cite{Gordon_Grempel}. 

From the above scaling, the SV-margin and the number of SV follow 
from the maximal stability $\kappa_{max}(0,\alpha)$ and the 
distribution of stabilities $\rho(0,\alpha ; \gamma)$ of the MSP 
in input space~\cite{Gordon_Grempel}. We obtain $\rho(k,\alpha;\gamma) = 
(\sqrt{2 / \pi}) \rho_1(k,\alpha) \, \Theta \left(\gamma - 
\kappa_{max}(k,\alpha)\right) \, + \,
\rho_0(k,\alpha) \, \delta\left(\gamma - \kappa_{max}(k,\alpha)\right)$ where 
$\rho_1(k,\alpha) = H[- \gamma/\tan\left(\pi \epsilon_g(k,\alpha)\right)] 
\exp(- \gamma^2/2)$ and $\rho_0(k,\alpha)$, the typical 
fraction of training patterns that belong to the SV, is such 
that $\rho(k,\alpha ; \gamma)$ integrates to one. For $\alpha \ll 1$, 
the SV-margin is $\kappa_{max}(k,\alpha) \sim \sqrt{(k+1)/\alpha}$ 
and $\rho_0(k,\alpha) \sim 1 - \sqrt{2 \alpha / \pi (k+1)} \exp{-(k+1)/2 \alpha}$, meaning that in that limit almost all the training 
patterns are SV. For $\alpha \rightarrow \infty$, 
$\kappa_{max}(k,\alpha) \sim 0.226 \sqrt{2\pi} (k+1)/\alpha$, and 
$\rho_0(k,\alpha) \sim 0.952 (k+1)/\alpha$, {\it i.e.} the 
typical number of SV is slightly smaller than the feature-space dimension. 
Solutions for other functions $\phi$ are more complicated, and we were not able to find a closed expression of $\epsilon_g(k,\alpha)$ for all $\alpha$. The function $\phi$ that gives the smallest generalization error at given $k$, at least for small $\alpha$, is $\phi(\lambda) = {\rm sign} (\lambda)$. But 
the fact that the generalization error increases with $k$ is 
a general property, independent of the function $\phi$.

We turn now to the more interesting problem of the capacity, defined as the typical number of dichotomies that the SVM may implement, a quantity closely related to the VC dimension of the learning machine~\cite{Opper}. We consider training sets where the patterns' classes  
are given by a random teacher, that selects outputs $+1$ and $-1$ with 
the same probability $1/2$. In this case, the order parameters are (\ref{order_param_b}) and (\ref{order_param_c}). The free energy is  
$f(k, \alpha, \kappa) = \max_{\tilde v_1,\tilde c_1,\tilde c_0} g(k, \alpha, \kappa; \tilde v_1,\tilde c_1,\tilde c_0)$ where $g(k, \alpha, \kappa; \tilde v_1,\tilde c_1,\tilde c_0)$ is obtained from (\ref{free_energy}) and (\ref{parameters}) by setting $R=0$.   

The capacity $\alpha_c(k)$, the largest reduced number of patterns that the machine can learn without errors, corresponds to a vanishing SV-margin, {\it i.e.} $\kappa_{\rm max}(k,\alpha_c(k)) = 0$. In this case, the extremae of $g(k, \alpha, 0; \tilde v_1,\tilde c_1,\tilde c_0)$ correspond to $\tilde c_0(\alpha,\kappa) = +\infty$ and $\tilde v_1 = \tilde c_1$ for all the possible functions $\phi$. This result means that the capacity is $\alpha_c = 2(k+1)$, independently of $\phi$, provided that the new features are uncorrelated. This result generalizes to other feature-spaces the value deduced  by Cover~\cite{Cover} through a geometrical approach that Mitchison and Durbin~\cite{Mitchison_Durbin} generalized to the case of quadratic separating surfaces. Notice that the latter corresponds to a SVM with $\phi(\lambda) = \lambda$ and $k = N$. The capacity of SVMs is smaller than the one of multilayered perceptrons with one hidden layer of $k+1$ neurons, which have the same number of degrees of freedom. For example, the capacity of the parity machine scales like $k \ln k$, and that of the committee 
machine like $k \sqrt{\ln k}$, for large $k$~\cite{Monasson_Zecchina,Kwon_Oh}. 

It turns out that in the case $\phi(\lambda) = {\rm sign} (\lambda)$, the 
maximal stability $\kappa_{\rm max}(k, \alpha)$ scales trivially with $k$. The 
order parameters are $\tilde v_1 = \tilde c_1 = 1$ so that 
$g(k, \alpha, \kappa; \tilde c_0)  =  g (0, \alpha/(k+1), \kappa; \tilde c_0)$, 
where the RHS corresponds to a SP of margin $\kappa$ in input space. 
The maximal stability is thus $\kappa_{\rm max}(k, \alpha) = \kappa_{\rm max}(0,\alpha/(k+1))$. From~\cite{Gordon_Grempel} we deduce that for 
$\alpha \ll 1$, $\kappa_{\rm max}(k, \alpha) \sim \sqrt{(k+1)/\alpha}$, 
and for $\alpha \rightarrow \alpha_c^-$, $\kappa_{\rm max}(k, \alpha) \sim \sqrt{\pi/8} \left( 2 (k+1)/\alpha - 1\right)$. If $\phi(\lambda) = \lambda$, 
the property $\kappa_{\rm max}(k,\alpha) \sim \kappa_{\rm max}(0,\alpha/k)$ 
is correct for $\alpha \ll k$. As $\kappa_{\rm max}(0,\alpha)$ is a concave decreasing function of $\alpha$~\cite{Gardner_Derrida}, including new features may result in a large  increase of the SV-margin. 

In most classification problems we expect that similar patterns belong to the same class. In that case, having a large SV-margin may be benefical for the generalization performance. In particular, if slightly corrupted versions of the training patterns are presented to the trained SVM, its output should not change. We consider a SVM that achieved error-free learning with a SV-margin $\kappa_{\rm max}>0$. We assume that the training patterns are corrupted, after the learning process, through ${\bbox \xi}^{\mu} \rightarrow  {\bbox \xi}^{\mu} + {\bbox \eta}^{\mu}$, where ${\bbox \eta}^{\mu}$ are randomly distributed vectors with probability distribution:

\begin{equation}
P({\bbox \eta}) = (2 \pi \Delta)^{-N/2} \exp \left( - {\bbox \eta}^2/2\Delta^2
 \right).
\end{equation}

\noindent We are interested in the classification error of the SVM on the corrupted patterns, defined as $\epsilon_t(k,\kappa,\Delta) = \sum_{\mu} (\sigma^{\mu}(\Delta) - \tau^{\mu})^2/(4P)$ where $\tau^{\mu}$ is the original pattern's class and $\sigma^{\mu}(\Delta)$ the SVM's output to the corrupted pattern. The dependance on $\alpha$ is implicitly included through $\kappa = \kappa_{\rm max}(k,\alpha)$. $\epsilon_t$
caracterizes the robustness of the SVM with respect to a small pattern's corruption ($\Delta \ll 1$). Input vectors close to a training pattern will be given its same class with probability $1-\epsilon_{t}(\kappa,\Delta)$. 

In the case of the linear-SVM (the SP), a straightforward calculation gives

\begin{equation} 
\epsilon_t(0,\kappa,\Delta) = H(-\kappa) H(\kappa/\Delta) + 
\int_{\kappa}^{+\infty} Dz H(z/\Delta).
\end{equation}

\noindent If the margin is $\kappa = 0$, one half of the training patterns have zero stability, and $\epsilon_t(0,0,\Delta) > 1/4$. Thus, any small perturbation results in misclassifications. If $\kappa > 0$, then $\epsilon_t(0,\kappa,\Delta) \sim \exp ( - \kappa^2/2\Delta^2 )$ for small $\Delta$. Consider next the general SVMs. If $\phi(\lambda) = {\rm sign} (\lambda)$ and $\kappa > 0$, $\epsilon_{t}(k,\kappa,\Delta) \sim \Delta$ for small $\Delta$. In comparison with the SP, the robustness of SVM is poor. This is due to the  discontinuity of the function $\phi$, as  a small perturbation of the input pattern may produce a strong perturbation on its stability. On the contrary, 
for continous functions $\phi$, like $\phi(\lambda) = 
\lambda$, and small $\Delta$, $\epsilon_{t}(k,\kappa,\Delta) \sim \exp ( - h(k) 
\kappa/\Delta )$ where $h(k)$ is an increasing function of $k$. 
Thus, continous functions $\phi$ are preferable for improving the SVM's robustness or noise tolerance. 

In conclusion, we presented the first study of the typical properties of a class of SVMs. We determined, as a function of the number of new features and the number of training patterns, the fraction of SV, the behaviour of the margin, the generalization error on a linearly separable task, the capacity and the probability of misclassification of training patterns slightly corrupted. Our results may explain why maximizing the margin is so important~: the probabilty that the trained SVM will assign the same class to the corrupted as to the original training patterns is enhanced by large margins.

\end{document}